# The Integration of On-Line Monitoring and Reconfiguration Functions using IEEE1149.4 Into a Safety Critical Automotive Electronic Control Unit

C. Jeffrey[1], R Cutajar[1], S Prosser[2], M Lickess[2], A. Richardson[1] & S Riches[3]

[1]Centre for Microsystems Engineering, Engineering Department, Lancaster University, Lancaster, LA1 4YR, UK

[2] TRW Automotive, Technical Centre, Stratford Road, Solihull, B90 4GW, UK

[3] Micro Circuit Engineering, Exning Road, Newmarket, Suffolk, CB8 0AU,UK

## Abstract

This paper presents an innovative application of IEEE 1149.4 and the Integrated Diagnostic Reconfiguration (IDR) as tools for the implementation of an embedded test solution for an Automotive Electronic Control Unit implemented as a fully integrated mixed signal system. The paper described how the test architecture can be used for fault avoidance with results from a hardware prototype presented. The paper concludes that fault avoidance can be integrated into mixed signal electronic systems to handle key failure modes.

## 1    Introduction

Today's motor vehicles contain an increasing number of microcontrollers (20 to 50), providing electronic control of a range of systems, including engine management, braking, steering and airbag safety systems. The demands of both the automotive market and physical environment puts pressure upon electronics design to ensure low cost, robust, high quality and high volume products. Quality targets of typically 10 parts per million are currently set by vehicle manufacturers. The trend to X-by-wire systems (e.g. braking and steering) and the proposed future electrical architectures are forcing the migration of the electronic control units to more harsh environments (e.g. brake caliper, engine, exhaust). Locating intelligence directly at the transducer interface is also becoming extremely desirable, as sub-system performance increases as signal noise is reduced, auto/intelligent calibration can be realised, connector count can be reduced and manufacturing costs optimised. It is also vital that both design and architectures meet the appropriate level of safety integrity and reliability. There is no doubt that future systems will be heavily dependent upon high-integrity, high-reliability embedded electronic modules.

The implications of these trends are a paradigm shift in automotive electronic design from board based to fully integrated systems. Table 1 provides some typical environmental specifications associated with the range of automotive applications.

Recently in [1] fault tolerant sensors and actuators for X-by-wire systems have been addressed. This paper highlights the technical difficulties associated with fault tolerant mechatronic systems but does not address the underlying electronics.

| Location | Typical Continuous Max Temperature | Vibration Level | Fluid Exposure |
| --- | --- | --- | --- |
| On Engine On Transmission | 140°C | Up to 15g | Harsh |
| At Engine (Intake Manifold) | 125°C | Up to 10g | Harsh |
| Under Hood (Near Engine) | 120°C | 3 - 5g | Harsh |
| Under Hood (Remote Location) | 105°C | 3 - 5g | Harsh |
| Exterior | 70°C | 3 - 5g | Harsh |
| Passenger Compartment | 70 - 80°C | 3 - 5g | Benign |

**Table 1 Electronic environmental conditions in automotive applications**

To address this, the work presented in this paper has led to the integration of test support hardware into an automotive ECU through an innovative application of IEEE1149.4 boundary scan [2] and IDR to address production test costs and test quality requirements, with further functionality including condition monitoring required to meet safety critical sub-system specifications. The paper is organised as follows. In section 2, the integration technologies required to realise low cost, safety critical ECU's are discussed. In section 3, the methods used to achieve auto health monitoring and fault avoidance are described. Section 4 describes the demonstrator and innovation in the application of the methods described in section 3. Section 5 draws conclusions.

## 2    Existing Strategies to Achieve Fault Avoidance

In the event of a potential failure, some kind of backup has to be provided to ensure certain reduced system functionality without complete system breakdown. In the majority of cases this may be achieved through circuit reconfiguration.

In many safety critical applications redundancy is used, in the form of both hardware and software, to ensure that the system will not fail [3, 4]. The subsystem would consist of duplicate devices that would run simultaneously; if one device becomes faulty the other device's output would be used instead. This technique is costly, as the entire functionality of the subsystem has to be duplicated. These systems are not economical for the automotive market place as it is impractical to implement an entire redundant subsystem. Automotive systems need to be produced cheaply and achieve high reliability. In failsafe systems additional circuitry will be required, the extent of which will depend on the architecture of the system and knowledge of the failure effects.

Having less than an entire redundant system available to achieve fault avoidance means that one will not be able to tolerate every possible fault. Instead, fault avoidance would be optimised to cope with faults that have significant impact on the functionality of the system. These faults need to be identified and ranked and this can be achieved through detailed Failure Mode and Effect Analysis (FMEA) and a Physics Of Failure (POF) approach. It is even possible to predict the type of failures that may occur in a newly designed system [5, 6]. Novel On-Line Monitoring (OLM) techniques have to be developed which detect these identified failure modes and also verify subsystem performance.

## 2.1 ECU Failure Analysis

Automotive electronics already exhibit a high degree of reliability within existing systems. This is currently achieved by using a combination of historical data, field returns and formal techniques (FMEA, failure rate analysis and fault tree analysis) to predict the areas of concern within the system. This is also supported by specific validating testing (temperature cycling, vibration etc) to focus upon items identified within the FMEA/failure analysis. Electronic component suppliers are required to submit specific validation tests results.

On reviewing detailed FMEA analysis of automotive Electric Control Unit's (ECU), we find that one of the most common failures are interconnect related on both discrete and IC devices. It is therefore imperative that:

- The number and density of interconnects are greatly reduced.
- There is a facility for interconnect testing at startup.
- There is a facility for on-line signal verification.

The lack of hardware related integrated on-line monitoring techniques for automotive systems means that there are few indications of the health and quality of a system in operation. When using components supplied from 3rd parties, test access to internal sensitive nodes of the device is not possible. It is therefore desirable to have a non-intrusive method to verify that system blocks are working correctly.

## 3 Health Monitoring and Reconfiguration Techniques Selected

### 3.1 System Monitoring

One of the key challenges in circuit fault avoidance is the variety of potential subsystem failure modes with varying failure intensity. Novel reconfiguration techniques may be dedicated to particular failure effects aiming at fault elimination though altering circuit topology. The results of an FMEA carried out on the subsystem will provide designers with failure modes that could occur in its operational lifetime. It is the faults arising from these failure modes that the subsystem needs to be configured around to ensure that the required level of operation is attained.

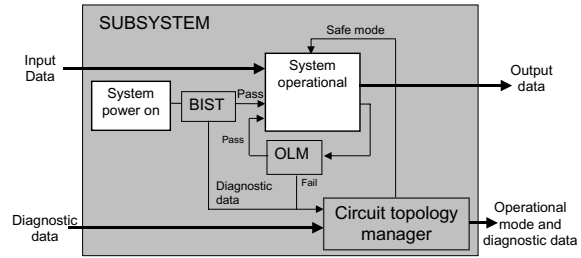

**Figure 1: Subsystem test and configuration flow**

In application, BIST methods could be used to check if failures are present on start up. If a failure is detected on a critical path, then the system would refuse to start as safety could be compromised. If the failure was on a non-critical path, then diagnostic data could be sent to a circuit topology manager, which would alter the circuit topology to provide its best performance at a required safety level. In this work, a circuit topology manager has been implemented and integrated into a novel system architecture as illustrated in (Figure 1). Once the system is operational, OLM techniques continuously check the subsystem for faults and notify the circuit topology manager as soon as a fault is detected.

### 3.2 Using IEEE1149.4 for on-line signal monitoring

IEEE 1149.1 Boundary Scan is a digital architecture that enables accessibility and observablity to circuit nodes and interconnect through a Test Access Port (TAP). The IEEE 1149.4 standard for a mixed-signal test bus [2] extends the test access structure for analogue stimulus injection and response evaluation via two buses connected to pins *AT1* and *AT2* respectively.

Recent research in 1149.4 [7,8] and its implementation and application [9, 10, 11] have shown that measurements for current, voltage level, frequency and phase are supported. However, measurement accuracy is limited due to capacitive and resistive characteristics of the Analogue Boundary Modules (ABM) that also cause degradation of the observed signal. Limitations in passive component measurements (resistance and capacitance) facilitating 1149.4 have been identified [12].

Building on previous review work [13] that suggested IEEE1149.4 could be used for on-line test, a circuit topology manager described in section 3.1 has been presented to control an 1149.4 standard architecture to implement both on-line monitoring and a circuit reconfiguration capability. (See Figure 2) The aim is to enable successive monitoring of circuit nodes with minimum area overhead, while avoiding an increase in pin count.

### 3.2.1 Circuit Topology Manger

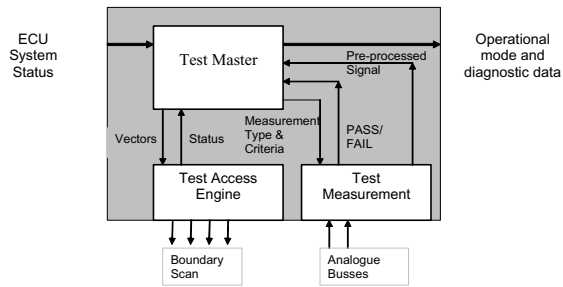

**Figure 2 Circuit Topology Manager**

To facilitate the IEEE1149.4 as an on-line monitoring technique an architecture is proposed for the Circuit Topology Manager in Figure 2.

*Test Master* This stores information required to execute a test. The test type and test data is passed to the *Test Measurement* circuit for simple tests e.g. DC voltage level (requires reference levels from *Test Master*) the test is evaluated directly and the results fed back to the *Test Master*. Once the *Test Measurement* circuitry is configured the vector information is passed to the *Test Access Engine* to connect *AT1* and *AT2* to the relevant circuit nodes.

*Test Access Engine* Converts the vectors containing boundary scan data from the Test Master and applies them to the ECU. When the system is correctly configured the engine informs the Test Master so the measurement can be performed.

*Test Measurement* This circuit is an analogue interface that conditions the signals on the analogue buses for processing/evaluation.

This architecture is based on sequential tests hence the timing data for each test node is also stored in the Test Master.

### 3.3 IDR

A sensing system designed using the IDR method [14] exploits existing multiple on-chip sensing that are switchable through electronic reconfiguration, reducing the need for redundant sensors used purely for test. A constraint to this technique is that the components within the system under test must exhibit a degree of replication to allow the connectivity of embedded functions to be changed. This takes advantage of the fact that many sensors exhibit some degree of component replication to remove cross-sensitivities and unwanted modes. During normal operation, these elements are dynamically re-arranged though the overall circuit function remains the same. When a fault occurs, one or more components cease to be equivalent to the others in the same group. When this component is interchanged with another, the overall circuit function changes. It has been shown that in the majority of cases, even minor parametric faults in analogue paths are easily observed at the system level, thus giving early fault detection. The profile of the changes is also observed, allowing the faulty component to be identified and therefore excluded. The circuit is then operational, but with reduced functionality in terms of signal or bandwidth[15].

The frequency at which the circuit reconfigures must be sufficiently different to the normal operational frequency to prevent IDR from interfering with the response time and efficiency. In this work IDR has been applied to a network of identical components providing visual and audible warnings to achieve tolerance to failure. This is further described in section 4.5.

## 4 Application to an X-By-Wire Control Unit

The specific objectives of this work are to establish a route to the realisation of the "zero failure" concept for automotive electronic systems by developing on-chip monitoring and self-test functions to detect in-field and production failures, and develop associated diagnostic algorithms to control system reconfiguration.

An Electronic Control Unit (ECU) was selected as the demonstrator for this work. This system combines a variety of power, signal and sensing components, all of which have to operate in a harsh environment. This type of ECU is designed to be mounted directly in the sensor environment, for example on engine or the brake calliper of the vehicle. The electronics can be subjected to a combination of temperature extremes, vibration, and a whole range of contaminants such as hydraulic oil, brake dust, and salts.

Due to the nature of the ECU demonstrator, a substrate technology which can withstand the harsh environment,

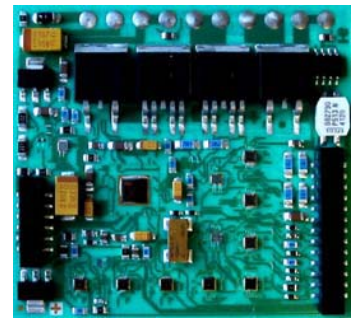

**Figure 3 IMS Electronic Control Unit**

dissipate heat efficiently and is compatible with hybrid assembly techniques had to be selected. The decision was made to use Insulated Metal Substrate (IMS) technology combined with bare die for the majority of the components and high temperature-resistant packaged components for the remaining devices. A two-layer IMS module was designed and manufactured on a copper base layer, with high temperature dielectric insulation between the metal layers. In order to avoid bumping the bare die components, standard die attach followed by aluminium wire bonding was carried out. Gold bonding was avoided to eliminate the formation of intermetallics at high temperatures. A nickel/gold surface finish was applied to the substrate, allowing Aluminium wire bonding and SMT component assembly. A high temperature resistant epoxy-based glob top material was also dispensed onto the bare die components as a protective layer. See **Figure 3**.

### 4.1 Add-On Board

In order to facilitate the circuit topology manager from Figure 2 a separate Add-On board was developed to emulate the automotive system to which the ECU was attached. Its primary functions are:

- Pass sensor data to the ECU.
- Allow the user to invoke test routines and induce faults.
- Act as the on-line test and reconfiguration engine.
- Convey information and regarding the ECU configured state.

### 4.2 Integrating Boundary Scan

The ECU was redesigned to use IEEE1149.1 compliant devices where possible, if an equivalent device was not found with boundary scan, then it was emulated via the placement of compliant devices in the signal path or in software. IEEE1149.4 compliance was implemented for the analogue nodes and the critical signal path. This enabled on-line probing of the critical signals in the system and in the event of a failure allows signals to be extracted, processed, bypassed and injected into the system.

Since 1149.4 is not fitted to devices as standard, the STA400 dual multiplexers from National Semiconductors were used as an add-on block to emulate compliance. Interfacing to the IC signal pins through hardwired STA400 devices, as illustrated in Figure 4.

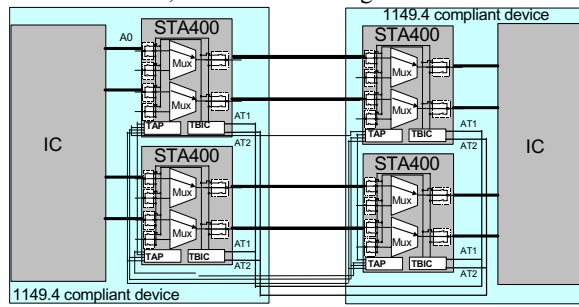

**Figure 4 Using STA400 devices as an emulation of 1149.4 compliant devices**

### 4.3 On-line Interconnect Monitoring

In order to monitor the integrity of an interconnect on-line, each side of a connection is linked to *AT1* and the other side is connected to *AT2*. The two buses are then fed into a differential amplifier to check to see if there is a difference in the signals. The output of the amplifier is then fed into a comparator circuit and is triggered to a logic high when there is a noticeable difference. See Figure 5.

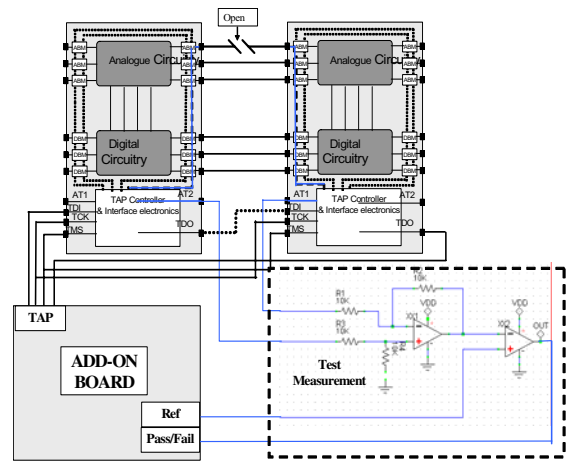

**Figure 5 On-line interconnect monitoring**

This test principle works on the basis that there is a signal present on the interconnect that one wishes to monitor.
Since ABMs are fitted into the critical signal path, even for the digital interconnects, it is possible to pass low amplitude signals ~ 0.5V to facilitate the detection of a connection for when the signals are at a logic low without placing an incorrect logic value in the circuit.

### 4.4 Motor Drive Circuitry

The existing H-Bridge circuitry was replaced with smart 'high side' BTS6510 and 'low side' BTS133 power drivers from Infineon and had an STA400 act as a transparent interface to emulate the drivers compliance to the standard. The advantages of using these devices are:

- The interconnect count would be greatly reduced.
- Would no longer require external test circuitry, as there are integrated voltage monitors and current monitoring circuits.
- An output pin on the high side driver provides diagnostic feedback to the microprocessor.

### 4.5 Lamps and Indicator Circuitry

The ECU had 3 lamps and a buzzer that were used as indicators for the operator. Even though their operation is not essential in the ECU's function, they are ranked as critical in the FMEA and failure as the status of the system cannot be conveyed to the user. This could result in incorrect operation of the vehicle e.g. a fault lamp cannot relay that extra caution is required from the user when a x-by-wire system is functioning incorrectly.
Since the indicator circuits are 4 identical circuits, it is possible to use the IDR technique to interchange them at a frequency above 85Hz. If there were faulty circuit/interconnect it would sequentially change what indicator it affected at a speed that was unnoticeable to the human eye. See Figure 6.

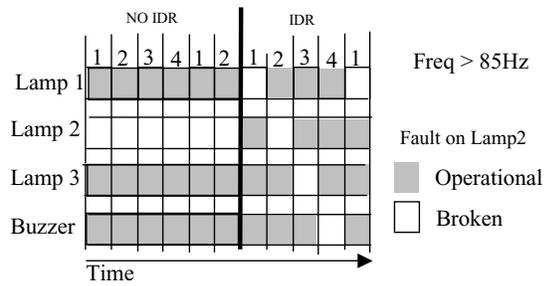

**Figure 6 Using IDR on Indicator Circuits**

## 5 Results

### 5.1 On-Line Monitoring of Circuit Nodes

| Signal | AT1 | AT2 | Detectable |
|---|---|---|---|
| Digital high (3.5v) | 3.5V | 0V | Yes |
| Digital low | 0.5V | 0V | Yes |
| Pull up | 3.5V | 0V | Intermittent |
| Pull down | Float | 0V | Intermittent |
| PWM | PWM | 0V | Yes |
| Analogue Ground | 0V | 0V | No |
| Hall sensor interface | Pulse train | 0V | Yes |

**Table 2 Types of signal lines and their detectability**

When monitoring digital signals, a bias of 0.5V was injected into the signal line in order to detect the connection when there was logic low. At high frequency of operation, propagation delays became apparent so a high frequency filter was added on the output of the differential amplifier to remove any peaks.

When monitoring analogue nodes a high impedance buffer is required as sensitive circuit nodes can easily be affected by this test method. It is also possible to measure signal characteristics through the 1149.4 architecture. The demonstrator uses Pulse Width Modulation (PWM) signals to control the speed of a motor. The frequency of the signal was 1kHz with a duty cycle of 60%. The Add-On board could detect changes of 0.001% in the duty cycle. When measuring DC values noise present in the system limited measurement accuracy to +/-10mV.

### 5.2 Signal Analysis

Results obtained on the limitations of the signal properties that can be measured are:

- **DC Response** : Results from passing a low frequency sinusoid through an ABM to identify maximum and minimum voltages to propagate through without clipping indicates that the maximum voltage that can be observed through the ABM is 3.92V. The minimum voltage observed was -640mV with an accuracy of +/- 10mV.
- **AC Response** : The 3dB cut-off frequency of the ABM has been determined through measurements:
  - *50kHz* stimulus frequency, no signal degradation in phase and magnitude can be observed.
  - *200kHz* a phase difference becomes observable. Hence one has to account for that difference in phase dependant measurements.
  - *1MHz* signals put through the ABM at this frequency show a 3dB cut-off.

### 5.3 Fault Avoidance

Once an interconnect failure is detected, the *AT1* and *AT2* buses are linked together to bypass the failure. This is possible on any nodes that are 1149.4 compliant; but once a fault is being compensated via *AT1* and *AT2*, no further monitoring can take place as the test system has its resources utilised. It is therefore important to segment the analogue bus lines into multiple functional groups so that multiple configurations can take place.

The IDR method works on the basis that the system configures the lamp and buzzer network constantly regardless of fault manifestation.

To verify that the ECU could function without the microprocessor, the power supply bond wires were removed. The Add-On board then monitored the switch circuitry until a transition of state was applied. Once this was detected, the Add-On board then accessed the circuit nodes that enabled the actuation of the motor drives. The correct signal was injected to operate the motor while the sensor status was monitored to know when to turn the motors off. Thus remote operation of the system was achieved. In addition, it has been verified that when an interconnect failure is injected into the system, fault free operation was observed. The technique is however, limited to coping with faults inside the reconfiguration matrix and should ideally be integrated into the silicon.

### 5.4 Limitations

The main challenge associated with the circuit topology manager is the detection and correction of a fault is not instantaneous. Therefore, careful analysis of what and when key parameters should be measured is required. The total time for conducting a test without repair is:

$$T_{total} = T_{con} + T_{Test}$$

Where $T_{con}$ is the time it takes to connect to the specific node. The worst case scenario would be to fully reconfigure each device in the scan chain between each test. The best case scenario it that the only 1 initial configuration is required to monitor all the relevant circuit nodes. This is shown in Figure 7.

The Term $T_{Test}$ is the time it takes to perform the measurement of the parameter. The 16 bit HCS12 processor used for implementation of the test master has an ADC capture time of 7µs which is very small in comparison to the configuration times above. However, computing a Fourier analysis of a sampled signal can be very time consuming to verify especially for very low frequency signals (~10Hz) and would result in an analysis time in the excess of 100ms. Using the above configuration and measurement times applied to a system where 10 circuit nodes are monitored and a test master with a 16MHz clock speed is used. The upper and lower bounds of the system performance are:

$$0.949\text{Hz} < T_{total} < 153\text{Hz}$$

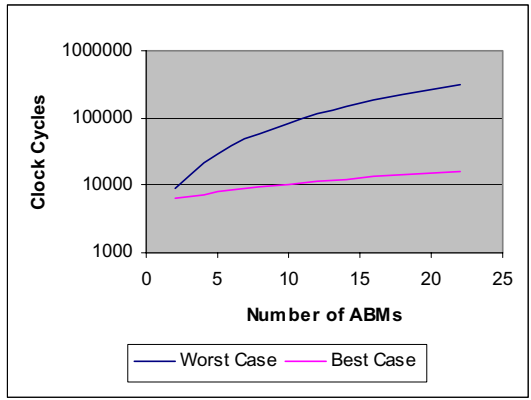

**Figure 7 Number of clock cycles Vs scan chain length**

With regard to fault tolerance the solution presented is limited to bypass a number of signals equal to half the number of analogue test buses that are implemented in the system. Once a fault is detected and compensated for, by using the reconfiguration technique presented, then subsequent testing will not be able to take place as the test resources are already in use. If there are 2 sets of buses then 1 bus set (*AT1* and *AT2*) can monitor sensors while the other bus set can perform actuations. This means that each circuit segment will still have only 1 bus set allocated to it so in the event of reconfiguration online test cannot continue in each segment. The amount of buses could be increased to create more circuit test segments but this will start to increase the complexity of the circuits and increase the potential for noise and crosstalk in the system.

## 6  Conclusion

The environments in which automotive systems are being placed are becoming increasingly harsh and the applications which they serve more safety critical. To achieve a high degree of reliability new packaging technologies need to be applied.
In this paper an innovative application of 1149.4 modules has been used to achieve a degree of fault avoidance in an automotive ECU that targets an X-by-wire application. The IDR concept has also been applied to manage the monitoring and reconfiguration strategies implemented in the system and proven in hardware. Results have verified that it is feasible to test, manage and reconfigure a fully integrated heterogeneous electronic systems on-line to achieve fault avoidance to key failure modes identified through an FMEA process.

### Acknowledgement

This work has been supported by EPSRC grant number GR/N36271/01 under the DREAM project.

The authors would like to thank National Semiconductor for access to their technology and prototype devices.